\documentclass[aps,prl,twocolumn,floats,showpacs,superscriptaddress]{revtex4}
\usepackage{graphicx,epsfig}
\usepackage{times}
\usepackage{graphics,dcolumn,bm,float}
\usepackage{amssymb,amsmath,rotate,color}

\begin{document}

\definecolor{red}{rgb}{1.0,0.0,0.0}
\definecolor{green}{rgb}{0.0,1.0,0.0}
\definecolor{blue}{rgb}{0.0,0.0,1.0}

\title{Josephson super-current in graphene-superconductor junction}

\author{E. Sarvestani}

\affiliation{Department of Physics, Isfahan University of
Technology, Isfahan 84156-83111, Iran}

\author{S. A. Jafari}
\affiliation{Department of Physics, Sharif University of Technology, Tehran 11155-9161, Iran}
\affiliation{School of Physics, Institute for Research in Fundamental Sciences (IPM), Tehran 19395-5531, Iran}

\begin{abstract}
Within the tunneling Hamiltonian formulation for the eight-component spinors,
the Josephson critical super-current has been calculated in a planar superconductor-normal 
graphene-superconductor junction. Coupling between superconductor regions and graphene is 
taken into account by a tunneling Hamiltonian which contains two types of tunneling, 
intra-valley and inter-valley tunneling. Within the present tunneling approach, we 
find that the contributions of two kinds of tunneling to the critical super-current, are 
completely separable. Therefore, it is possible to consider the effect of the inter-valley 
tunnelings in the critical super-current. The incorporation of these type of processes 
into the tunneling Hamiltonian, exposes a special feature of the graphene Josephson 
junctions. The effect of inter-valley tunneling appears in the length dependence 
plot of critical current in the form of oscillations. We also present the results 
for temperature dependence of critical super-current and compare with experimental 
results and other theoretical calculations.   
\end{abstract}


\maketitle
\section{Introduction}
In the past few years, after the first experimental synthesis by 
Geim et.al~\cite{geim} the community has witnessed the growing interest in mono-layer carbon 
atoms arranged in a honeycomb lattice, called graphene. The peculiarity of graphene 
electronic structure is the primary reason which has given rise to such enthusiasm among scientists. 
Linear dispersion relation near the discrete Fermi points and chiral nature of carriers 
in graphene are responsible for the most intriguing phenomena that have been 
reported~\cite{graphene1,graphene2}. 

Among the attractive properties of graphene, from both basic and applied point of view, 
are the superconductor proximity effect which has been studied experimentally and 
theoretically. Before any experimental observation, Beenakker~\cite{prl2006} predicted 
in a seminal work that at the graphene-superconductor interface, not only the 
conventional Andreev retro-reflection takes place, but also it can exhibit specular 
Andreev reflection owing to presence of two different valleys in the graphene band 
structure. Specular Andreev reflection takes place when an electron in the conduction 
band converts into a hole in the valance band. At the low doping level, specular reflections 
are dominated and at Dirac point all reflections are of this type~\cite{beenakker2008}. 
By solving the Dirac-Bogoliubov-de Gennes (DBdGD) equations for an ``ideal'' 
Normal-Superconductor (NS) interface, in the short junction limit, in which coherence 
length $ \xi = \hbar v_F/ \Delta_{\circ}$ ($v_F $ is the Fermi velocity and $\Delta_{0}$ 
the superconductivity order parameter), is much larger than junction length, $L$, 
and with neglecting inter-valley scatterings, Titov and Beenakker~\cite{titov} and 
simultaneously Moghaddam and Zareyan~\cite{malek} calculated the Andreev bound states, 
with which they obtained Josephson super-current. Their results predicted the existence 
of a finite current even at the Dirac point. In these works, to solve the DBdG equations, 
the rigid boundary condition was assumed, according to which superconductivity gap has 
a fixed and finite value in the superconductor regions and it is zero in the normal region. 

First experimental investigation of superconductor-graphene-superconductor (SGS) 
Josephson junction~\cite{nature2007} showed that Josephson current does flow through 
these junctions. This current depends strongly on the position of Fermi energy and as 
it had been predicted, there was a non-zero super-current at the neutral Dirac point. 
Afterwards, several experimental studies have been 
conducted~\cite{experiment1,experiment2,experiment3,experiment4,experiment5}. Recently, 
in addition to critical super-current, direct measurement of the current-phase 
relation (CPR) has been performed with the interferometry technique~\cite{mason,solidi}. 
The results confirm the predicted non-sinusoidal CPR curves ~\cite{titov} and show 
that the deviation from the sinusoidal behavior increases linearly with critical current. 

At the same time, more theoretical efforts were triggered~\cite{mason,blachschaffer1,blachschaffer2,blachschaffer3,gonzalez1,gonzalez2,hagymasi,jadid1}. Self-consistent 
solution of the tight binding DBdG equation, at zero 
temperature~\cite{blachschaffer1,blachschaffer2} and later generalization to finite 
temperature~\cite{blachschaffer3}, have provided results on the dependence of the 
Josephson super-current on length, temperature, doping level, phase difference and 
pairing symmetries of the superconductivity order parameter. Another method based on 
Cooper pair propagation over long distances for junctions with $ L \gg W$ also reveals 
such dependencies in the graphene Josephson junctions~\cite{gonzalez1,gonzalez2}. 
Shifting the Fermi energy away from zero point results in an enhancement in the 
critical super-current due to finite density of states at the Fermi level for $\mu\neq 0$. 
Besides the enhancement, one can see oscillations in the diagrams of the critical 
super-current as a function of length~\cite{gonzalez1, gonzalez2,hagymasi}.

In this paper we will use the perturbative Green's function method in framework of 
path integral and tunneling Hamiltonian between superconductor and normal graphene 
regions to calculate the critical Josephson super-current. An s-wave superconductivity 
pairing will be assumed in the superconductor areas of the graphene honeycomb lattice. 
In contrast to previous works, we don't neglect the inter-valley processes,
as such processes may become important in the presence of edges, or external
potentials with sharp variations~\cite{adam}.
Among the other circumstances where inter-valley scattering becomes important,
one can mention the presence of Kekule textures~\cite{Mudry} which may arise
from creation of instantons of opposite sign in two valleys~\cite{Vaezi}.
We will incorporate both types of inter-valley and intra-valley tunneling processes into 
the Hamiltonian. We will discuss the temperature and length dependence of the critical 
super-current. The advantage of the present tunneling approach is that, within this approach, 
contributions of inter-valley and intra-valley processes to the total Josephson super-current 
can be totally separated. Therefore, the effect of the inter-valley tunneling on the critical 
current can be isolated. The present formulation reveals a peculiar feature of the graphene 
Josephson junctions which is the result of incorporating the inter-valley tunnelings. 
In the curve of critical super-current as a function of junction length, the effect of 
inter-valley tunneling appears in the form of oscillations which are drastically different 
from oscillations due to non-zero chemical potential.

The paper is organized as follows. First, we summarize the perturbative Green's function 
method~\cite{mori} and rephrase it for the 8-component spinors needed in the present paper.
We then apply the formulation to the problem of SGS junctions. We close the paper by
a discussion on the results.

\section{Method and Model}
 In a Josephson junction between two superconductors, a dissipation-less 
 super-current flows from one superconductor to the other to provide a state with
 minimum energy~\cite{josephson1}. The critical super-current is given by,
 \begin{equation}
I= \frac{2e}{\hbar} \frac{\partial F}{\partial \varphi} \equiv I_{c}  \sin(\varphi)\, ,
 \label{super-current}
 \end{equation}
where $ F $ is the free energy of the system, $I_c$ indicates the critical super-current 
and $\varphi=\varphi_R - \varphi_L $ is phase difference between two superconductors.
In the framework of the path integral, partition function is given by,
\begin{equation}
  Z =  e^{-\beta F} = 
\int D[ \bar{\Psi} ] D[ \Psi ] e^{-S( \bar{\Psi} , \Psi)},
\label{partition}
\end{equation}
in which $S$ is the effective action of system that is a function of Grassmann 
variables $\Psi$ and $\bar{\Psi}$. In the imaginary time formalism ($\tau=it$), 
the action will be given by,
\begin{equation}
S = \sum_{\vec{k}} \int_{0}^{\beta} d\tau [ \bar{\Psi} (\vec{k}, \tau) \partial_{\tau} \Psi (\vec{k}, \tau) + H(\bar{\Psi} (\vec{k}, \tau), \Psi (\vec{k}, \tau)) ] .
\end{equation}  
Here, $H$ is the full Hamiltonian of the system that takes the form
\begin{equation}
\hat{H}=\hat{H}_{L}+\hat{H}_{R}+\hat{H}_{N}+\hat{H}_{T} \, .
\end{equation}
$\hat{H}_{L} , \hat{H}_{R} , \hat{H}_{N} $ and $\hat{H}_{T}$ are respectively the 
Hamiltonian of the left and right  superconductors, normal region and tunneling between 
superconductors and normal regions. The form of these Hamiltonians can be quite general. 
So that the formula~\eqref{freeenergy} holds under very general circumstances. 
The particular form of these Hamiltonians suitable for our own problem are given in 
Eqs. \eqref{SC}, \eqref{tunnel1}, \eqref{tunnel2}. If we express the matrix form of the 
above Hamiltonian as sum of two parts, $\hat{T}$ as tunneling term and $\hat{g}^{-1}_0$ 
for the rest, then in the basis 
$ \hat{\Psi} (\vec{k}) \equiv ( \hat{\Psi} (\vec{k}_{L}), \hat{\Psi} (\vec{k}_{N}), \hat{\Psi} (\vec{k}_{R}) )\, $, 
it can be written in the compact matrix form as,
\begin{equation}
\hat{H}=\sum_{\lbrace \vec{k} \rbrace} \hat{\Psi} ^{\dagger} (\vec{k})  [ \hat{g}_{0}^{-1} + \hat{T} ]   \hat{\Psi} (\vec{k}) \,.
\end{equation}
where $\hat{g}^{-1}_0$ and $\hat{T}$ are given by the following matrix forms:
\begin{equation}
\hat{g}_{0}^{-1} = \left( \begin{array} {ccc}
g_{0 L}^{-1} & 0 & 0 \\
0 & g_{0 N}^{-1} & 0 \\
0 & 0 & g_{0 R}^{-1}  \\
\end{array}
\right) \, ,
\end{equation}
\begin{equation}
\hat{T} = \left( \begin{array} {ccc}
0 & T_{NL} & 0 \\
T_{NL}^{\ast} & 0 & T_{NR}^{\ast}\\
0 &  T_{NR} & 0 \\
\end{array}
\right) \, .
\end{equation}
Here, $g_{0L}^{-1}$, $g_{0N}^{-1}$ and $g_{0R}^{-1}$ are respectively the inverse propagators
for the left, normal and right areas excluding the tunneling parts.

After a Fourier transform, the effective action becomes,
\begin{equation}
S = \sum_{{\vec{k} , i\omega_{n}}} \bar{\Psi} (\vec{k} , i\omega_{n}) [-i\omega_{n} \mathbb{I} + \hat{g}_{0}^{-1} + \hat{T} ] \Psi (\vec{k} , i\omega_{n}) \, .
\end{equation}
where $\hbar \omega_{n} = \pi k_{B} T (2n+1)$ are the Matsubara frequencies. With definition, 
\begin{align}
-\hat{G}_{0}^{-1} &= -i\omega_n \mathbb{I} + \hat{g}_{0}^{-1}\, ,
\end{align}
and from Eq.~\eqref{partition} we have,
\begin{equation}
e^{-\beta F} = \int D [\bar{\Psi}] D [\Psi] e^{ - \sum_{\vec{k} , i\omega_n} \bar{\Psi} (\vec{k}, i\omega_n) [-\hat{G}_{0}^{-1} + \hat{T} ] \Psi (\vec{k} , i\omega_n) } .
\end{equation}
This is a Gaussian integral, which can be performed to lead to the following 
equation for the free energy,
\begin{align}
F &= -\frac{1}{\beta} {\rm Tr} \ln [-\hat{G}_{0}^{-1} + \hat{T} ] \nonumber \, , \\
&= -\frac{1}{\beta} {\rm Tr} [\ln (-\hat{G}_{0}^{-1}) + \ln (1-\hat{G}_{0} \hat{T})]\, .
\end{align}
Here, Tr means a summation over all diagonal matrix elements and an integration over 
all momenta. Having expanded this expression in terms of the tunneling amplitude, it 
can be readily shown that the fourth-order term is the first non-zero and leading term that 
contributes to the Josephson super-current. Using Eq.~\eqref{super-current} and the identity 
$\ln \det \hat{A}={\rm Tr} \ln \hat{A}$, the final expression for super-current 
is obtained as,
\begin{align}
I (\varphi)&=\frac{-2e}{\hbar} \frac{1}{4\beta}\frac{\partial}{\partial \varphi} {\rm Tr}(\hat{G}_{0} \hat{T})^4  \label{freeenergy} \\
&=\frac{-2e}{\hbar \beta} \frac{\partial}{\partial \varphi}{\rm Tr} [ \hat{G}_L \hat{T}_{NL} \hat{G}_N \hat{T}_{NR}^{\ast} \hat{G}_R \hat{T}_{NR} \hat{G}_N \hat{T}_{NL}^{\ast} ]\nonumber .
\end{align}

Now let us specialize the general formula~\eqref{freeenergy} to the case special
case of SGS junctions of this paper.
\begin{figure}
\centering
\includegraphics*[width=8.5cm]{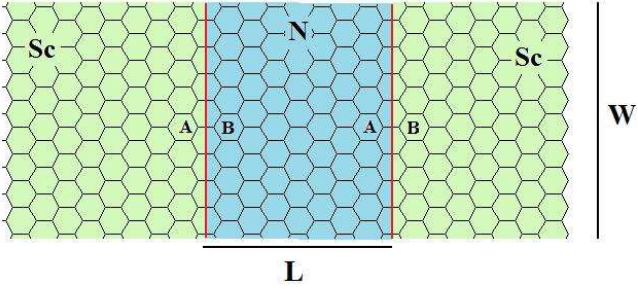}
\caption{\small{The schematic geometry of graphene Josephson junction. $L$ is the separation of two superconductor (Sc) regions and $W$ is the junction width.}}
\label{josephson}
\end{figure}
 The geometry of the graphene Josephson junction that we consider in this paper is depicted 
in Fig.~\ref{josephson}. In the normal graphene region, the low energy electrons are governed 
by the relativistic Dirac Hamiltonian, 
$ H_N = v_F (\vec{k}_N\cdot\vec{\sigma})\otimes \tau_{0}$,
where $v_F\simeq10^6 m/s$ is the Fermi velocity in graphene and $\vec{\sigma}$ and $\tau$ 
are Pauli matrices. $ \vec{\sigma} $ operates on the sub-lattice degrees of freedom 
and $ \tau_{0} $ acts on the valley degree of freedom. Dirac fermions in graphene 
live in two different valleys, $K$ and $\bar{K}$, on the opposite corners of the Brillouin zone.

Pristine graphene cannot be superconductor in ordinary conditions. Experimentally, to prepare 
an SGS junction, superconducting electrodes can be deposited on top of a graphene sheet, 
so that due to this proximity of superconductor electrodes to graphene layer, 
the superconductivity order parameter can be assumed to be induced on the left and right
graphene regions. Here, an s-wave superconductivity  will be assumed in the left and right 
superconducting areas. In our model, the assumption is that the superconducting pairing 
takes place between two time-reversal electrons on the same sub-lattice but on different 
valleys in the Brillouin zone ~\cite{prb2010}. Thus the mean field Hamiltonian for the 
superconducting part will be,
\begin{equation}
\hat{H}^\alpha_{\rm SC}=\sum_{\vec{k},\sigma} \Delta_{0} e^{i\varphi_\alpha} a_{\vec{k} , \sigma}^{\dagger} \bar{a}_{-\vec{k}, -\sigma}^{\dagger} +\Delta_{0} e^{i\varphi_\alpha} b_{\vec{k} , \sigma}^{\dagger} \bar{b}_{-\vec{k} , -\sigma}^{\dagger} + h.c,
\label{SC}
\end{equation}
where $a$ and $b$ are fermion operators on sub-lattices A and B pertaining to valley $K$, 
while $\bar{a}$ and $\bar{b}$ denote the corresponding operators on valley $\bar{K}$. 
$\Delta_{0}$ is the magnitude of superconducting order parameter and $\varphi_\alpha$ 
denotes phase of each $\alpha=R,L$ superconducting leads.

The electron tunneling between superconductors and normal graphene region can be 
considered in several different ways. In our case, as it is depicted in Fig.~\ref{josephson}, 
superconductor and normal graphene  are assumed to be connected to each other 
through their zigzag edge. Tunneling Hamiltonian contains left and right tunneling between 
superconductor and normal graphene. As can be seen in Fig.~\ref{josephson}, for right side, 
electron tunneling takes place between sub-lattice A in normal region and sub-lattice B 
in superconductor region while at the left interface, electrons tunnel between sub-lattice 
B in normal region and sub-lattice A in superconductor area. Therefore, the tunneling 
Hamiltonian in tight binding approximation can be written as,
\begin{equation}
\hat{H}_{T}=\gamma_T \sum_i \hat{B}_{iR}^{\dagger} \hat{A}_{iN} + {A}_{iL}^{\dagger} \hat{B}_{iN} + h.c \, ,
\label{tunnel1}
\end{equation}
where $\gamma_T$ is the tunneling amplitude and $\hat{A}$ and $\hat{B}$ generally stand 
for the Dirac fermion operators on the A and B sub-lattices and summation extends over 
all atomic sites along the interfaces. By writing the Dirac fermion operators as the sum 
of two valley operators, tunneling Hamiltonian for the right side will be given by,
\begin{align}
 H_T^R&=\gamma_T \sum_{k_N,k_R} e^{-i(k_{Nx}-k_{Rx})x_R} ( b_{k_R}^{\dagger} a_{k_N} + e^{i \vec{Q}\cdot \vec{R}_R} \bar{b}_{k_R}^{\dagger} a_{k_N}  +
\nonumber \\
& \qquad e^{-i \vec{Q}\cdot \vec{R}_R} b_{k_R}^{\dagger} \bar{a}_{k_N} + \bar{b}_{k_R}^{\dagger} \bar{a}_{k_N} ) + h.c ,
\label{tunnel2}
\end{align}
and likewise for the left interface. As can be seen, tunneling Hamiltonian contains 
two types of tunneling, intra-valley (first and last terms) and inter-valley 
(second and third terms) tunnelings. Inter-valley tunneling involves a "momentum transfer" 
$\vec{Q}$, that connects the two independent valleys in the Brillouin zone.
Now,  let us define an appropriate basis to rewrite all parts of Hamiltonian. 
A suitable representation for Fermi operators is the following:
\begin{equation}
\mathbf{\Psi} (\vec{k}) = (a_{k\uparrow}^{\dagger},b_{k\uparrow}^{\dagger},\bar{a}_{-k\downarrow},  \bar{b}_{-k\downarrow},\bar{a}_{k\uparrow}^{\dagger},\bar{b}_{k\uparrow}^{\dagger},a_{-k\downarrow},b_{-k\downarrow})^T  .
\end{equation} 
In this basis, $\hat{G}_{0 N}^{-1}$ and $\hat{G}_{0 \alpha}^{-1}$ ($\alpha = R,L $) 
are block diagonal and one can easily obtain the required matrices by just inverting 
the blocks. The Green's function of the normal region is written as,
 \begin{equation}
 \hat{G}_{0 N}^{-1} = i\omega_n \mathbb{I} + \hbar v_{F} \left(
 \begin{array}{cccc}
 \vec{\sigma}\cdot \vec{k}_{N \uparrow}&0&0&0 \\
 0& \vec{\sigma}\cdot \vec{k}_{N \downarrow}&0&0\\
 0&0& \vec{\sigma}^{\ast}\cdot \vec{k}_{N \uparrow}&0\\
 0&0&0& \vec{\sigma}^{\ast}\cdot \vec{k}_{N \downarrow}
 \end{array} \right) .
 \end{equation}
It is important to mention that in the normal area two electrons propagate between 
two superconductors as a Cooper pair with different spin. Thus, $\vec{k}_{N \uparrow}$ 
and $\vec{k}_{N \downarrow}$ are independent degrees of freedom in this region. 
For the superconductor Green's function we have,
\begin{align}
\hat{G}_{0 \alpha}^{-1} = i\omega_n \mathbb{I}_{8 \times 8} + \bigg[ \mathbb{I}_{2\times 2} \otimes ( \Delta e^{i \varphi_{\alpha} \sigma_z} \sigma_x) \bigg] \otimes \mathbb{I}_{2\times 2} + \nonumber \\
\qquad \left(
\begin{array}{cc}
-(\hbar v_F \vec{\sigma} \cdot \vec{k}_{\alpha}) \otimes \sigma_z & 0 \\
0 & (\hbar v_F \vec{\sigma}^{\ast} \cdot \vec{k}_{\alpha}) \otimes \sigma_z
\end{array} \right)  .
\end{align}
Finally, for tunneling part of the Hamiltonian we have, 
\begin{equation}
T_{N\alpha} = \bigg[ \left( 
\begin{array}{cc}
\delta_{\alpha}^{\uparrow} & 0 \\
0 & -\delta_{\alpha}^{\downarrow}
\end{array} \right) \otimes (e^{i \theta_{\alpha} \sigma_z} \sigma_x) \bigg] \otimes \mathbb{I}_{\alpha},
\end{equation}
where we use these notations:
\begin{align}
 \mathbb{I}_L &= \dfrac{1}{2}\left(\sigma_x+i \sigma_y\right) = \left(
 \begin{array}{cc}
 0&1\\
 0&0
 \end{array} \right)  , \\
\mathbb{I}_R &= \dfrac{1}{2}\left(\sigma_x-i \sigma_y\right) = \left(
 \begin{array}{cc}
 0&0\\
 1&0
 \end{array} \right)  ,\\
 \delta_{\alpha}^{\sigma}&=\gamma_T e^{i(k_{Nx}^{\sigma}-k_{\alpha x})x_{\alpha}},~~~~~ 
  e^{i\theta_{\alpha}}=e^{i \vec{Q}\cdot \vec{R}_{\alpha}} .
\end{align}
Now, we have all of matrices needed for calculation of the free energy. 
From Eq.~\eqref{freeenergy} and by using these matrices, we obtain:
\begin{widetext}
\begin{equation}
F=-\dfrac{16\Delta^2 \gamma_T^4(\hbar v_F)^2}{\beta}  \dfrac{\big[ (1+\cos\theta ) k_x^{\downarrow}k_x^{\uparrow} + (-1+\cos\theta) k_y^{\downarrow}k_y^{\uparrow} - \sin\theta (k_x^{\downarrow}k_y^{\uparrow}+k_x^{\uparrow}k_y^{\downarrow})\big] \cos(\varphi+ (k_{x}^{\downarrow}-k_{x}^{\uparrow})L)} {D_L D_R [(\hbar \omega_n)^2+(\hbar v_F)^2 k_{\downarrow}^2][(\hbar \omega_n)^2+(\hbar v_F)^2 k_{\uparrow}^2]}  \, ,
\label{sub}
\end{equation}
\end{widetext}
where $\vec{k}$ is a collective index denoting all possible momenta, i.e. 
$\vec{k} = \{\vec{k}_L,\vec{k}_R,\vec{k}_{N}^{\uparrow},\vec{k}_{N}^{\downarrow}\}$. 
In Eq.~\eqref{sub}, we have used the following notations,
 \begin{align}
\theta &= \theta_L - \theta_R = \vec{Q}\cdot \vec{L} \nonumber \, , \\ 
D_\alpha&=(\hbar \omega_n)^4+(\hbar v_F)^4 k_\alpha^4+\Delta_{0}^4+ \nonumber \\
& \qquad 2((\hbar \omega_n)^2 k_\alpha^2+(\hbar \omega_n)^2 \Delta_{0}^2+\Delta_{0}^2 k_\alpha^2) \nonumber \, ,\\ 
E_\alpha^2&=(\hbar \omega_n)^2+\Delta_{0}^2+(\hbar v_F)^2 k_\alpha^2 .
\end{align}
The summation in Eq.~\eqref{sub} is over all Matsubara frequencies and all momenta around 
the pertinent valleys. In the limit of wide junction limit, i.e. $W \gg L$, the details
of the interface becomes irrelevant and we can replace the summation with an integration 
around the corresponding valley, up-to an energy cutoff that the linear dispersion holds.

Due to symmetry of the integration region, even terms of integrand can contribute 
to the free energy. With definition~\eqref{super-current} for critical super-current 
and from Eq.~\eqref{sub} we obtain (see Appendix),
\begin{equation}
I_c = \dfrac{128ek_BT\Delta^2(T)\gamma_T^4 S^2 (1+\cos\vec{Q}\cdot \vec{L})}{\hbar \pi^2} \sum_{\omega_n}\big[ f(i\omega_n) g(i\omega_n)\big]^2 \, 
\label{current}
\end{equation}
where $S=WL$ is the area of junction and functions $f$ and $g$ are given by,
\begin{widetext}
\begin{align}
f(i\omega_n)&=\dfrac{1}{(\hbar v_F)^2}\int_0^{E_c}\dfrac{x((\hbar \omega)^2+\Delta_0^2(T)+x^2) dx}{x^4+2((\hbar \omega_n)^2+\Delta_0^2(T))x^2+(\hbar \omega_n)^4+\Delta_0^4(T)+2\Delta_0^2(T) (\hbar \omega_n)^2}  ,\\
g(i\omega_n)&=\dfrac{1}{(\hbar v_F)^2}\int_0^{E_c}\dfrac{x \sin(x) dx}{\sqrt{x^2+(\hbar \omega_n)^2}} \tan^{-1}\sqrt{\dfrac{1-x^2}{x^2+(\hbar \omega_n)^2}} .
\label{current1}
\end{align}
\end{widetext}
$E_c$ is the cutoff energy and $\vec{L}$ indicates the vector connecting 
two superconducting leads.

Carefully tracing the above derivation, shows that in Eq.~\eqref{current}, all effects 
of the inter-valley tunnelling processes appears in a term proportional to 
$\cos(\vec{Q} \cdot \vec{L})$. 
The elimination of the term proportional to $\sin(\vec{Q} \cdot \vec{L})$ can be 
physically understood: In view of symmetry of the problem under $y \rightarrow -y$ 
transformation, the Hamiltonian is symmetric under $k_y \rightarrow -k_y$, which implies 
that free energy must be symmetric with respect to $\vec{Q} \rightarrow -\vec{Q}$ 
transformation. Hence only the term proportional to $\cos(\vec{Q} \cdot \vec{L})$ 
will survive in the free energy. Therefore, an oscillatory behavior in the critical 
super-current is predicted which is solely due to inter-valley processes. 
It can be seen in the final result by setting $\vec Q=0$, which amounts to ignoring 
the inter-valley processes, $(1+\cos(\vec Q.\vec L))\rightarrow g_v=2$.
This is the simple valley degeneracy expected in all extensive quantities, 
when the inter-valley processes are ignored~\cite{adam}. 
\begin{figure}[t]
\centering
\includegraphics[width=8.5cm]{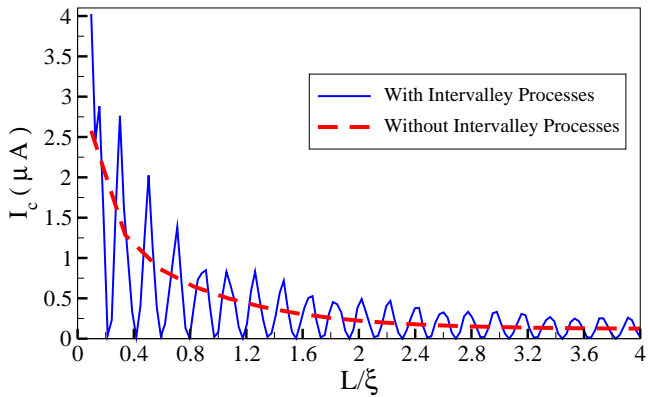}
\caption{\small{Plot of the critical super-current as a function of junction length for $T=0.1T_c$. The solid line with sharp oscillations shows the total critical current and the dashed line presents the contribution of the intra-valley tunneling processes to the critical super-current.}}
\label{pic1}
\end{figure}

The fact that contribution
of inter-valley processes (both in tunneling and superconducting pairing) gives rise
to an additive modulation by a wave vector $\vec Q$ is a feature of the present
tunneling formulation. 
This shows that within the present formulation the contributions of two 
types of tunneling to the critical super-current are naturally separable,
and reproduces to proper degeneracy factor $g_v=2$ in the limit where inter-valley
processes are ignored. The oscillatory factors of $1+\cos(\vec Q.\vec r)$ arise
in other contexts, which require a convolution of Green's functions related to
Dirac cones~\cite{Sherafati}.
\section{Results and Discussion}
To calculate the critical super-current from Eq.\eqref{current} we need to 
have the temperature dependence of the magnitude of superconductivity order
parameter, $\Delta(T)$. For s-wave superconductors the pair 
potential satisfies~\cite{abrikosov},
\begin{equation}
\ln \dfrac{\Delta(T=0)}{\Delta(T)}=2\sum_{n=1}^{\infty}(-1)^{n+1} K_{0}(\dfrac{n\Delta(T)}{k_BT})\, ,
\label{dama}
\end{equation}
where $K_{0}$ is the modified Bessel function and $ \Delta_{0}(T=0) $ is the magnitude 
of the superconductivity gap at zero temperature. Analytical calculation of Eq.~\eqref{current} 
is not feasible, thus we will calculate it numerically. Electron dispersion relation in 
graphene remains linear below a momentum cutoff that corresponds to 
$E_c = \hbar v_F k_c \approx 1eV$. Careful scrutiny of the Eq.~\eqref{current1} 
reveals an $\omega_n^{-6}$ dependence of the summand, so that it would rapidly decay at 
larger Matsubara frequencies. Therefore, it is adequate to sum over only first few frequencies.
We are now ready to present the results of calculations of the critical current using 
Eq.~\eqref{current}. As indicated before, we are in the wide junction limit $(W \gg L)$, 
but we have no restriction on the distance between the superconductors, i.e we can 
calculate the critical current for all range of the distances ($L < \xi$ , $L \simeq \xi$ 
and $L > \xi$). In all of our calculations we set $W=10 \mu m \gg \xi $.
\begin{figure}[t]
\centering
\includegraphics*[width=8.5cm]{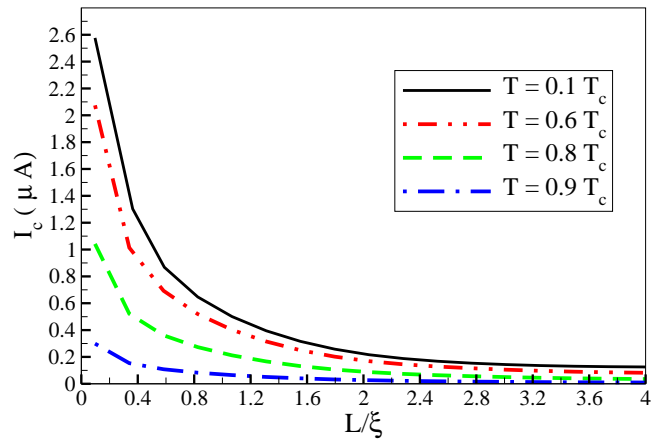}
\caption{\small{The critical super-current as a function of junction length for 
different temperatures. For easier comparison, in this plot, the inter-valley 
tunneling processes have not been taken into account. Decreasing the critical 
super-current by increasing junction length and temperature, is clearly seen 
in these curves. }}
\label{pic2}
\end{figure}

Figure~\ref{pic1} shows the length dependence of the critical current. The temperature 
at which the calculation has been performed is $0.1 T_c$. In this figure, the curve with 
sharp oscillations shows the total critical current, while the smooth digram gives the 
contribution of the intra-valley tunneling processes to the critical current. 
As it is obviously expected, the current decreases by increasing the distance between two 
superconductors. Although the decreasing trend of two curves is similar to each other, 
but the oscillatory behavior of the total critical current discriminates between two 
types of tunneling. In fact, the distinguishing feature of the inter-valley electron 
tunneling is the oscillation with a characteristic wave-vector 
${\vert \vec{Q} \vert}$ in the total current. 
This suggests that if one measures the critical current versus the junction 
length, an oscillatory behavior is expected which can be attributed to the 
inter-valley tunnelings.

The presence of a second valley connected to the first one by time reversal symmetry, 
is the reason for emergence of a novel specular Andreev reflection~\cite{prl2006,beenakker2008}. 
The present tunneling formulation reveals that existence of two valleys leads to another 
peculiar feature in Josephson current through graphene, namely oscillatory behavior of the
Josephson current as a function of the junction length, $L$. In fact, our results in 
Fig.~\ref{pic1} shows that for some distance between two superconducting electrodes, 
the critical super-current will be strongly suppressed due to inter-valley tunnelings 
between normal and superconducting regions. This can be physically interpreted as the
destructive interference between the super-current arising from Andreev bound states~\cite{malek}, 
and those arising from Andreev modes~\cite{prl2006} (made possible by specular reflections). 
Note that, in all our calculations, an undoped normal graphene region has been assumed, 
so that, undoubtedly the doping ~\cite{gonzalez1,gonzalez2,hagymasi} is not responsible 
for the oscillatory behavior of the critical super-current.
\begin{figure}[t]
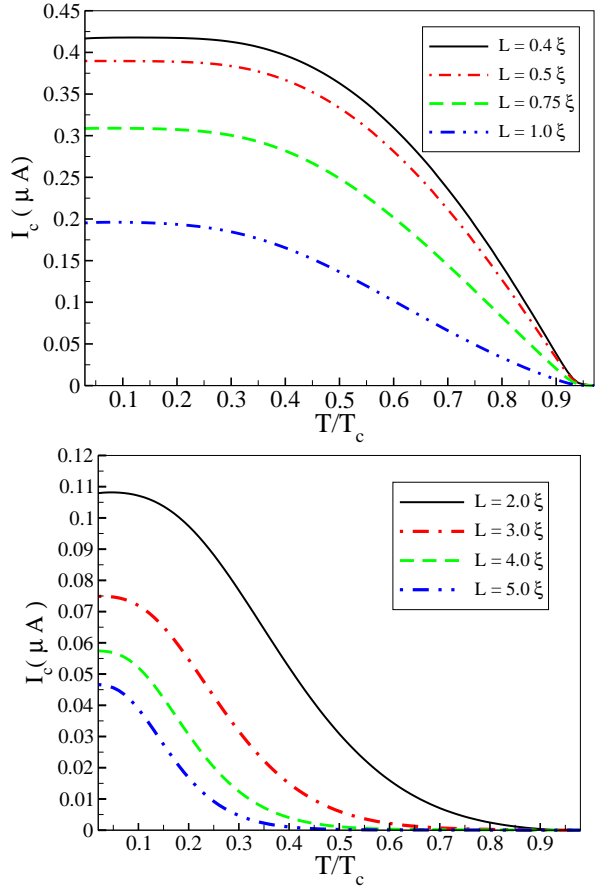

\centering
\includegraphics*[scale=0.33]{Fig4-a-pic3.eps}
\includegraphics*[scale=0.33]{Fig4-b-pic4.eps}
\caption{\small{Temperature dependence of the critical super-current for different junction length (up) for $L<\xi$ and (down) $L>\xi$. Notice that in these diagrams only the contribution of the intra-valley tunnelings to the critical current, have been taken into account. }}
\label{pic3}
\end{figure}

Figure~\ref{pic2} presents the length dependence of critical super-current for different 
temperatures. In this plot, to show the general dependence of the critical current
on the junction length more clearly, and also to facilitate comparison with other works
which do not consider inter-valley tunneling, we have only retained the intra-valley tunnelings.
It is clearly seen that the critical super-current suddenly decreases in all diagrams, 
when one approaches $L \approx \xi$ from below. Furthermore, regarding the temperature
dependence, it drops very rapidly when temperature approaches the critical temperature. 
Our results are in a good agreement with the results of the self-consistent 
tight-binding method (Fig.~5a in Ref.~\cite{blachschaffer1} and
Fig.~11a in Ref.~\cite{blachschaffer2})

In Figure~\ref{pic3} we have plotted the contribution of the intra-valley processes 
in the critical super-current as a function of temperature. The results are displayed 
separately for distances below the coherence length, $ L < \xi $ (top panel) and distances 
above it, $ L > \xi $ (bottom panel). The top panel of this figure agrees with results of
Ref.~\cite{gonzalez2}. 
When one compares the temperature dependence for junctions lengths less than the
coherence length, with those above the coherence length, a clear qualitative difference 
can be observed: For distances below the coherence length, there is a plateau in 
super-current for temperatures smaller than $0.5 T_c$ and after that all diagrams 
reach zero near to $T=T_c$. But for distances above the coherence length, after a 
much shorter plateau, an exponential decay takes place for temperatures near $T_c$. 
As one can see in the Fig.~\ref{pic3}, the temperature at which rapid decrease sets in, 
decreases by increasing the junction length.
\begin{figure}
\centering
\includegraphics[scale=0.32]{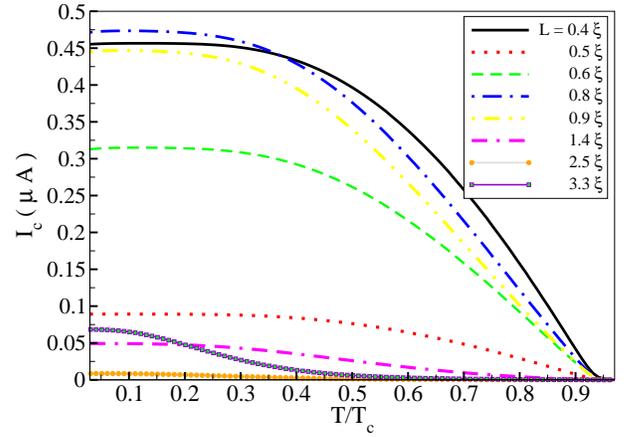}
\caption{\small{Plot of the total critical super-current as a function of 
temperature for different values of the distance between superconducting electrodes. 
In contrast to the diagrams of the Fig. \ref{pic3}, in this diagram the 
contributions of all processes have been calculated.}}
\label{pic5}
\end{figure}

Temperature dependence of the critical current has been measured in several research works. 
Our results are qualitatively consistent with these experimental measurements. In particular, 
when a comparison between the observations in Ref.~\cite{experiment3} and Ref.~\cite{experiment5} is made, 
the above-mentioned  different behaviors of temperature dependence of the critical 
current can be noticed. Therefore from our calculations we judge that in 
samples of Ref.~\cite{experiment3}, the junction length is smaller than the coherence length, 
while for those in Ref.~\cite{experiment5}, the length of the junctions seems to be above the 
coherence length.
Similar temperature dependencies have been obtained in the other theoretical works. 
For example, in Fig.~2 of Ref.~\cite{blachschaffer3} and Fig.~3 of Ref.~\cite{gonzalez2}, 
for $ L < \xi $ and $ L > \xi $, similar results have been obtained. 
Finally, in Fig.~\ref{pic5} the total critical super-current as a function of temperature 
has been plotted. In this diagram both inter-valley and intra-valley tunnelings have been 
taken into account. As can be seen in this figure, the variation of critical 
current with the junction length is different from diagrams of Fig.~\ref{pic3},
which is a consequence of the oscillatory behavior the critical current 
as a function of the junction length. Otherwise, the
general trends are similar to those in Fig.~\ref{pic3}.

In summary, we employed a method based on perturbative Green's function in the framework 
of path integral, to calculate the critical super-current in the graphene Josephson junction. 
Our results presented the length and temperature dependence of the critical super-current. 
The role of the inter-valley tunnelings in the super-current has been investigated and 
it is shown that incorporating these kind of tunnelings, led to sharp oscillations in the
behavior of critical current as a function of the junction length. Comparison of our 
results with the results that have been obtained by the other formalisms and also 
with experimental observations, suggests that the junctions realized in experiment 
are likely to be in weak tunneling regime which are consistent with perturbative treatment.

\section{Acknowledgement}
We thank F. Shahbazi and A. Vaezi for insightful comments and discussions.
S.A.J. was supported by the National Elite Foundation (NEF) of Iran.

\appendix
\section{Details of calculations}
To obtain Eq.~\eqref{current} from Eq.~\eqref{sub}, we should expand the numerator of 
Eq.~\ref{sub}. For convenient, we replace 
$k_x^{\downarrow} L, k_x^{\uparrow} L, k_y^{\downarrow} L$ and $k_y^{\uparrow} L$ 
respectively with $x , x^{\prime} , y$ and $y^{\prime}$. There are two main terms: 
For the one proportional to $\sin\varphi$, we have,
\begin{align}
\sin\varphi & \bigg\{ \bigg[(1+\cos\theta)xx^{\prime} + (-1+\cos\theta) yy^{\prime} -\sin\theta (xy^{\prime} + x^{\prime}y) \bigg] \nonumber \\  & \big[ \sin x \cos x^{\prime} - \cos x \sin x^{\prime}\big] \bigg\} \, .
\end{align}
These terms are odd functions at least with respect to one of their variables while 
the integration region is even respect to all variables, hence they do not have contribution 
to the free energy. The other term is proportional to $\cos\varphi$,
\begin{align}
\cos\varphi & \bigg\{ \bigg[(1+\cos\theta)xx^{\prime} + (-1+\cos\theta) yy^{\prime} -\sin\theta (xy^{\prime} + x^{\prime}y) \bigg] \nonumber \\ & \big[ \cos x \cos x^{\prime} + \sin x \sin x^{\prime}\big] \bigg\} \, .
\end{align}
After careful consideration we find out that only the term proportional 
to $x\,x^{\prime}\,\sin x \, \sin x^{\prime}$ has non-zero contribution to the free energy.
Therefore, the remained integration corresponding to normal region variables is of this type:
\begin{equation}
\int_{-E_c}^{E_c} \, dy \, \int_{-\sqrt{E_c^2-y^2}}^{\sqrt{E_c^2-y^2}} dx \, \dfrac{x \sin x}{(\hbar \omega_n)^2 + x^2 + y^2} \, .
\end{equation} 
Integration of variable $y$ can be performed analytically, which gives,
\begin{equation}
2 \,\, \int_{0}^{E_c} dx\, \dfrac{x\,\sin x}{\sqrt{(\hbar \omega_n)^2 + x^2}} \, \arctan \sqrt{\dfrac{E_c^2 - x^2}{(\hbar \omega_n)^2 + x^2}} \, .
\end{equation}


\begin{thebibliography}{50}

\bibitem{geim}
K. S. Novoselov, A. K. Geim, S. V. Morozov, D. Jiang, Y. Zhang, S. V. Dubonos, 
I. V. Grigorieva and A. A. Firsov, Science, \textbf{306}, 666 (2004).
\bibitem{graphene1}
A. K. Geim and K. S. Novoselov, Nature Mater. \textbf{6},183 (2007). 
\bibitem{graphene2}
A. H. Castro Neto, F. Guinea, N. M. R. Peres, K. S. Novoselov, A. K. Geim, 
Rev. Mod. Phys. \textbf{81},109 (2009).
\bibitem{prl2006}
C. W. J. Beenakker, Phys. Rev. Lett. \textbf{97}, 067007 (2006).
\bibitem{beenakker2008}
C. W. J. Beenakker, Rev. Mod. Phys. \textbf{80}, 1337 (2008).
\bibitem{titov}
M. Titov and C. W. J. Beenakker, Phys. Rev. B \textbf{74}, 041401 (2006).
\bibitem{malek}
A. G. Moghaddam and M. Zareyan, Phys. Rev. B \textbf{74}, 241403 (2006).
\bibitem{nature2007}
H. B. Heersche, P. Jarillo-Herrero, J. B. Oostinga, L. M. K. Vandersypen and A. F. Morpurgo, 
Nature \textbf{446}, 56 (2007).
\bibitem{experiment1}
H. B. Heersche, P. Jarillo-Herrero, J. B. Oostinga, L. M. K. Vandersypen and A. F. Morpurgo, 
Eur. Phys. J. Spec. Top. \textbf{148}, 27 (2007).
\bibitem{experiment2}
A. Shailos, W. Nativel, A. Kasumov, C. Collet, M. Ferrier, S. Guéron, R. Deblock and H. Bouchiat, 
Eur. Phys. Lett. \textbf{79}, 57008 (2007).
\bibitem{experiment3}
X. Du, I. Skachko and E. Y. Andrei, Phys. Rev. B \textbf{77}, 184507 (2008).
\bibitem{experiment4}
C. M. Ojeda-Aristizabal, M. Ferrier, S. Gueron and H. Bouchiat, 
Phys. Rev. B \textbf{76}, 165436 (2009).
\bibitem{experiment5}
D. Jeong, J. Choi, G. Lee, S. Jo, Y. Doh and H. Lee, Phys. Rev. B \textbf{83}, 094503 (2011). 
\bibitem{solidi}
C. Girit, V. Bouchiat, O. Naaman, Y. Zhang, M. F. Crommie, A. Zettl, I. Siddiqi, 
Phys. Status Solidi B \textbf{246}, 2568 (2009).
\bibitem{mason}
C. Chialvo, I. C. Moraru, D. J. Van Harlingen, N. Mason, arXiv:1005.2630 (unpublished).
\bibitem{blachschaffer1}
A. M. Black-Schaffer, S. Doniach, Phys. Rev. B \textbf{78}, 024504 (2008).
\bibitem{blachschaffer2}
J. Linder, A. M. Black-Schaffer, T. Yokoyama, S. Doniach, A. Sudbo, 
Phys. Rev. B \textbf{80}, 094522 (2009).
\bibitem{blachschaffer3}
A. M. Black-Schaffer, J. Linder, Phys. Rev. B \textbf{82}, 184522 (2010).
\bibitem{gonzalez1}
J. González, E. Perfetto, Phys. Rev. B \textbf{76}, 155404 (2007).
\bibitem{gonzalez2}
J. González, E. Perfetto, Journal of Physics Condensed Matter \textbf{20}, 145218 (2008).  
\bibitem{hagymasi}
I. Hagymási, A. Kormányos, J. Cserti, Phys. Rev. B \textbf{82}, 134516 (2010).  
\bibitem{jadid1}
Q. Sun and X. C. Xie, Journal of Physics Condensed Matter \textbf{21}, 344204 (2009).
\bibitem{adam}
S. Das Sarma, S. Adam, E. Hwang, E. Rossi, Rev. Mod. Phys. {\bf 83}, 407 (2011).
\bibitem{Mudry} Chang-Yu Hou, C. Chamon, C. Mudry, Phys. Rev. Lett. {\bf 98}, 186809 (2007).
\bibitem{Vaezi} A. Vaezi, Xia-Gang Wen, arxiv:1101.1662.
\bibitem{mori}
M. Mori, S. Hikino, S. Takahashi, S. Maekawa, 
Journal of the Physical Society of Japan \textbf{76}, 054705 (2007).
\bibitem{josephson1}
B. D. Josephson, Phys.Lett. \textbf{1}, 251 (1962)
\bibitem{prb2010}
N. B. Kopnin and E. B. Sonin, Phys. Rev. B \textbf{82}, 014516 (2010).
\bibitem{Sherafati} 
S. Saremi, Phys. Rev. B \textbf{78} 184430 (2007); 
M. Sherafati, S. Satpathy, Phys. Rev. B \textbf{83} 165425 (2011).



\bibitem{abrikosov}
A. A. Abrikosov, L. P. Gorkov and I. Y. Dzyaloshinskii, Quantum Field Theoretical Methods in Statistical Physics, 2nd ed. (Pergamon Press, Oxford, London, UK, 1965).

\end{thebibliography}
\end{document}